\documentclass[prb,twocolumn,nofootinbib,floatfix]{revtex4-1}
\usepackage{color}
\usepackage{xcolor}
\usepackage{graphicx}
\usepackage{bm}
\usepackage{mathrsfs}
\usepackage{amsmath}
\usepackage{amsthm}
\usepackage{amssymb}
\usepackage{amsfonts}
\usepackage{bbold}

\usepackage{ifpdf}
\ifpdf
   \usepackage[unicode=true,colorlinks=true]{hyperref}
\fi

\newcommand{\KP}{${\bm k} \cdot{ \bm p}$}

\newcommand{\ket}[1]{\left\vert #1 \right\rangle}

\begin{document}

\title{Optical orientation and alignment of excitons in ensembles of inorganic perovskite nanocrystals}

\author{M.O.~Nestoklon$^{1,*}$}
\author{S.V.~Goupalov$^{1,2,\dag}$} 
\author{R.I.~Dzhioev$^1$}
\author{O.S.~Ken$^1$}
\author{V.L.~Korenev$^1$}
\author{Yu.G.~Kusrayev$^1$}
\author{V.F.~Sapega$^1$}
\author{C.~de~Weerd$^3$}
\author{L.~Gomez$^3$}
\author{T.~Gregorkiewicz$^3$}
\author{Junhao Lin$^4$}
\author{Kazutomo Suenaga$^4$}
\author{Yasufumi Fujiwara$^5$}
\author{L.B.~Matyushkin$^6$}
\author{I.N.~Yassievich$^1$}

\affiliation{$^1$~Ioffe Institute, 194021 St. Petersburg, Russia\\
$^2$~Department of Physics, Jackson State University, Jackson MS 39217, USA\\
$^3$~Van der Waals-Zeeman Institute, University of Amsterdam, Science Park 904, 1098 XH Amsterdam, The Netherlands\\
$^4$~National Institute of Advanced Industrial Science and Technology (AIST), AIST Central 5, Tsukuba 305-8565, Japan\\
$^5$~Division of Materials and Manufacturing Science, Graduate School of Engineering, Osaka University, 2-1 Yamadaoka, Suita, Osaka 565-0871, Japan\\
$^6$~St. Petersburg Electrotechnical University LETI, 197376 St. Petersburg, Russia\\
$^*$Email:nestoklon@coherent.ioffe.ru; $^{\dag}$Email:serguei.goupalov@jsums.edu}

\begin{abstract}
  We demonstrate the optical orientation and alignment of 
  excitons in a
  two-dimensional layer of CsPbI\textsubscript{3} perovskite 
  nanocrystals prepared by colloidal synthesis 
  and measure the anisotropic exchange 
  splitting of exciton levels in the nanocrystals.
  From the experimental data at low temperature (2~K), we obtain the average
  value of anisotropic splitting of bright exciton states
  of the order of 120~$\mu$eV.
  Our calculations demonstrate that there is a significant
  contribution to the splitting due to
  the nanocrystal shape anisotropy for all inorganic perovskite 
  nanocrystrals.
\end{abstract}

\maketitle

\section{Introduction}

Hybrid organic-inorganic halide perovskites are new optoelectronic
materials that have attracted enormous attention as solution-deposited
absorbing layers in solar cells with power conversion efficiencies of
above 20\%.\cite{Green14,Chien14}
The excellent optoelectronic properties of 
organic halide perovskite thin films 
are comparable to those of conventional direct-gap semiconductors 
like GaAs, which makes the perovskites promising for 
generating and detecting spin.\cite{Zhang15} 
The potential of halide perovskites for spintronic applications 
is just beginning to be investigated.\cite{Odenthal17}

While all-inorganic Cs--Pb-halide perovskite 
nanocrystals (NCs), fabricated via colloidal synthesis, 
have been introduced only recently,\cite{Protesescu15}
they have already greatly impressed the scientific community 
by their unprecedented brightness and abnormally 
high photoemission rates. 
This shows good promise for a range of applications
from color-converting phosphors and 
light-emitting diodes\cite{Tan14}
to lasers.\cite{Xing14,Yakunin15}
Various spectroscopic techniques 
and theoretical considerations have been applied to conduct a 
comprehensive study of spectral and dynamical characteristics of 
single- and multi-exciton states in CsPbX\textsubscript{3}
NCs with X being either Br, I, or 
their mixture.\cite{Makarov16,Lin16,Weerd16,Jong17,Becker18}

The fine structure of exciton levels in CsPbBr$_3$\cite{Fu17} and
CsPbI$_3$\cite{Yin17} NCs has been probed by means of the single-dot spectroscopy. 
Splittings of the bright exciton level into linearly polarized components were detected which 
were on the order of $1$~meV for CsPbBr$_3$ NCs and several hundred $\mu$eV for CsPbI$_3$ NCs
with the mean size of $\sim$ $10$~nm. However, Yin {\it et al.} emphasized that, for some samples, the fine structure splittings for all NCs were below their spectral resolution of $200$~$\mu$eV, although the same synthesis procedure was nominally adopted.\cite{Yin17} Switching of the NC emission from the split exciton line to a single red-shifted trion line was also observed in these experiments 
when the photoexcitation power was above a certain threshold.\cite{Fu17,Yin17}
Recently ensemble measurements of magneto-photoluminescence have been undertaken for CsPbBr$_3$
NCs\citep{Canneson17} and 90\% of photoluminescence was attributed to negatively charged trions, though excitation power density did not exceed $1$~W/cm$^2$.\cite{Canneson17}

In this work we investigate optical orientation and alignment of excitons in 
ensembles of CsPbI\textsubscript{3} NCs.
Our experimental approach has been previously 
used to probe the fine structure of excitons in bulk 
III-V and II-VI semiconductors\cite{bookPikus,bookOO} and 
heterostructures\cite{Ivchenko95,bookIvchenkoPikus} 
with quantum wells,
quantum dots,\cite{Dzhioev97, Dzhioev97_PRB,Dzhioev98_1,Dzhioev98_2}
and quantum wires.\cite{Dzhioev03}
This technique allows one to measure a splitting of bright exciton 
levels averaged over an ensemble. 
This technique is not limited 
by the spectral resolution and allows one to measure splittings as low as few $\mu$eV.
It also allows to study the ensembles \textit{in situ} with account on possible 
interaction between the NCs.\cite{Lin16}

In our samples containing a two-dimensional layer of CsPbI\textsubscript{3}
NCs, the suppression of the optical orientation at
zero magnetic field, together with the strong
optical alignment of photoexcitations, indicates that photoluminescence (PL) is dominated by the neutral excitons.
The dependence of the degree of the photoluminescence linear polarization on the magnetic field  
allowed us to measure the ensemble-averaged splitting between linearly polarized components of the
bright exciton state of 120~$\mu$eV.

The paper is organized as follows. In Sec.~\ref{sec:bandstructure} we 
briefly review what is known on
the crystal structure and the electronic 
band structure of inorganic perovskites.
A comprehensive theoretical analysis of the origin of the bright exciton
level splitting is presented in Sec.~\ref{sec:exchange}. We also estimate 
the splitting caused by the NCs shape anisotropy from the \KP\ analysis.
The details of sample synthesis and experimental measurements are presented in 
Secs.~\ref{sec:synthesis} and \ref{sec:PL}, respectively.
In Sec.~\ref{sec:results} we discuss experimental results which enable us to 
extract the value of the ensemble-averaged splitting between linearly
polarized components of the bright exciton state. 
In Sec.~\ref{sec:discussion} we discuss the origin of this splitting.
In Sec.~\ref{sec:conclusions} the conclusions are drawn.

\section{Bandstructure of inorganic halide perovskites}\label{sec:bandstructure}

The inorganic perovskites CsPbI\textsubscript{3}, 
as well as other bulk APbX\textsubscript{3} perovskite materials, 
are known to form at least three
different phases:\cite{Moller58,Fujii74,Fu17}
the high-temperature cubic phase ($O_h$ point group), 
the tetragonal phase ($D_{4h}$), 
and the low-temperature orthorhombic phase ($D_{2h}$). 
There exist reports of yet another low-temperature, monoclinic  phase.\cite{Fujii74}
All phase transitions in the bulk perovskites occur well
above room temperature.\cite{Moller58,Fujii74}
In ultrathin two-dimensional CsPbBr\textsubscript{3} halide perovskites,
the coexistence of different phases at room temperature has been observed.\cite{Alivisatos16}
For NCs, the phase transition temperatures may be 
strongly shifted by the presence of the surface\cite{Haruyama16,Murali16,Wang17}
and the synthesis method proposed in Ref.~\onlinecite{Protesescu15} leads to 
all CsPbX$_3$ NCs crystallization in the cubic phase.\cite{Protesescu15}
Fu {\it et al.}\cite{Fu17} and Yin {\it et al.}\cite{Yin17}, basing on X-ray 
diffraction measurements, also reported cubic crystal structure of CsPbBr$_3$ 
and CsPbI$_3$ NCs, respectively, at room temperature.
However, recent comparison of the X-ray diffraction spectra of CsPbBr$_3$ NCs with the calculated pair distribution function allowed Cottingham and Brutchey\cite{cottingham16} to conclude that the room-temperature crystal structure of these NCs is orthorhombic.

A detailed analysis of the perovskites symmetry can be found 
in the literature.\cite{Even15,Even15lett,Yu16}
Most of the basic properties of inorganic halide perovskites may be understood from an analysis of the cubic 
phase. In this phase crystals have the point group $O_h$ 
coinciding with the group of the wave vector at the $R$ point of the Brillouin zone, where the band extrema are located.
The conduction and valence bands of lead halide perovskites arise from the cationic $p$ and $s$ orbitals, respectively, {\it i.e.} the band ordering is reversed as
compared to classical semiconductors. 
Near the $R$ point the structure of the conduction band can be described by the effective Hamiltonian
\begin{equation}
\hat{H}({\bf k})=\hat{H}_0({\bf k})+\hat{H}_{SO} \,,
\end{equation}
where
\begin{multline}
\hat{H}_0({\bf k})=-A \, k^2 + 3 \, B \sum\limits_{\alpha} \hat{J}_{\alpha}^2 
\left( k_{\alpha}^2 - {k^2}/{3} \right)
\\
 +2 \, \sqrt{3} \, D \, \sum\limits_{\alpha>\beta} \left\{ \hat{J}_{\alpha} \hat{J}_{\beta} \right\}_s \, k_{\alpha} \, k_{\beta} \,,
\end{multline}
\begin{equation}
\hat{H}_{SO}=\frac{\Delta}{3} \, \sum_{\alpha} \hat{J}_{\alpha} \hat{\sigma}_{\alpha} \,,
\end{equation}
$A$, $B$, and $D$ are the conduction band parameters, $\alpha,\beta=x,y,z$, $\hat{J}_{\alpha}$ are the matrices of the angular momentum $j=1$, $\hat{\sigma}_{\alpha}$ are the Pauli matrices, $\Delta$ is the spin-orbit splitting of the conduction band, and $\left\{ \hat{J}_{\alpha} \hat{J}_{\beta} \right\}_s=\left( \hat{J}_{\alpha} \hat{J}_{\beta} + \hat{J}_{\beta} \hat{J}_{\alpha} \right)/2$. 

The band structure of the low symmetry phases 
may be considered as the folded band structure of the 
cubic phase, with the small effect of symmetry reduction.\cite{Even15}
The electronic states in the vicinity of the band gap are 
folded from the $R$ point of the Brillouin zone onto the $\Gamma$ point 
and the crystal field splitting is added.

The tetragonal ($D_{4h}$) phase is characterized by the crystal field splitting described by
\begin{equation}
\hat{H}_{CF}^{(1)}= \epsilon_1 \, \left[ \hat{J_z}^2 - {2}/{3} \right] \,,
\end{equation}
while the orthorhombic ($D_{2h}$) phase is associated with additional crystal field splitting of the
form
\begin{equation}
\hat{H}_{CF}^{(2)}= \epsilon_2 \, \left[ \hat{J_y}^2 -  \hat{J_x}^2 \right] \,.
\end{equation}

For the $D_{4h}$ case ($\epsilon_2=0$), the resulting Hamiltonian can be diagonalized at $k=0$.
In particular, this yields the basis Bloch wave functions for the irreducible representation
$\Gamma_6^-$ of the group $D_{4h}$ which corresponds to the lowest conduction band.
They are
\begin{align}
|\Gamma_6^-, \uparrow \rangle= \frac{\cos{\xi}}{\sqrt{2}} \, (X+i \, Y) \, \downarrow +
\sin{\xi} \, Z \, \uparrow \,,
\label{Bloch1}
\\
|\Gamma_6^-, \downarrow \rangle= \frac{\cos{\xi}}{\sqrt{2}} \, (X-i \, Y) \, \uparrow -
\sin{\xi} \, Z \, \downarrow \,,
\label{Bloch2}
\end{align}
where $\tan{2 \xi}=\frac{2 \sqrt{2} \Delta}{\Delta - 3 \, \epsilon_1}$.
The phases of these functions are chosen to yield the basis functions of the representation
$\Gamma_7$ of the group $T_d$ (analogous to the representation $\Gamma_6^-$ of the group $O_h$) in the limit $\epsilon_1=0$.\cite{bookIvchenko} These functions, along with the valence-band wave functions,
\begin{align}
|\Gamma_6^+, \uparrow \rangle= S \, \uparrow \,,
\\
|\Gamma_6^+, \downarrow \rangle= S \, \downarrow \,,
\end{align}
will be used in Sec.~\ref{subsection:lr}
to obtain the matrix element of the 
long-range electron-hole exchange interaction.

\section{Fine structure of exciton levels}\label{sec:exchange}
\begin{figure}[tbp]
  \centering{
   $
   \begin{matrix}
   \ket{0,0} = & \phantom{-}  {\color{blue}\uparrow}{\color{red}\downarrow}-\phantom{i} {\color{blue}\downarrow}{\color{red}\uparrow} \\ \hline
   \ket{1,x} = & - {\color{blue}\uparrow}{\color{red}\uparrow} - i {\color{blue}\downarrow}{\color{red}\downarrow} \\
   \ket{1,y} = & \phantom{-}   {\color{blue}\uparrow}{\color{red}\uparrow} - i {\color{blue}\downarrow}{\color{red}\downarrow}  \\
   \ket{1,z} = & \phantom{-}  {\color{blue}\uparrow}{\color{red}\downarrow}+ \phantom{i} {\color{blue}\downarrow}{\color{red}\uparrow}    \end{matrix}
   $
   \raisebox{-0.5\height}{\includegraphics{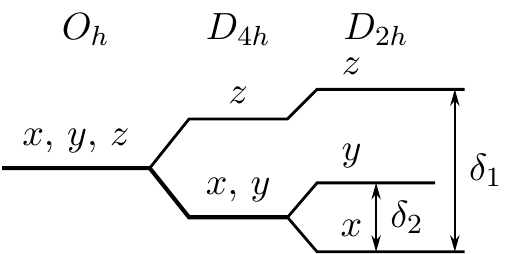}}
  }
  \caption{Qualitative scheme of the exchange splitting of 
  bright exciton triplet state $\vert 1, xyz \rangle$ 
  for different symmetries of the NC. 
The actual order of split levels may be different. } 
\label{fig:levels_scheme}
\end{figure}

For an exciton in a crystal with point group $O_h$,
formed by the electron from the $\Gamma_6^-$ band and the hole from
the $\Gamma_6^+$ band,
the isotropic part of the electron-hole exchange interaction splits the 
fourfold degenerate exciton level into an optically inactive singlet of 
$\Gamma_1^-$ symmetry $\ket{0,0}$ and optically active triplet of $\Gamma_4^-$
symmetry $\ket{1,xyz}$ (Fig.~\ref{fig:levels_scheme}). 
A reduction of symmetry due to anisotropy of NC shape or 
lowering of the crystal symmetry to $D_{4h}$
allows for the splitting of the bright exciton level 
into the ``in-plane'' doublet and 
the $\Gamma_2^-$ optically active singlet. For $D_{2h}$ point symmetry, 
all excitonic levels are non-degenerate.

In a NC of highly anisotropic shape, even when the crystal structure 
is cubic, the symmetry allows for the full splitting of 
the bright excitonic level into the $\ket{x}$, $\ket{y}$, and $\ket{z}$
states.
Below we use optical orientation measurements to 
extract information about the fine structure of 
the bright exciton states
in NCs of CsPbI\textsubscript{3}. 
The fine structure of the bright exciton is defined by the 
exchange interaction between the electron and the hole. 
Microscopically, the exchange splitting 
may be considered as the sum of two contributions: 
the long-range (non-analytic) and the short-range (analytic).\cite{BirPikus,Denisov73}
In NCs, the former contribution is sensitive to
the shape of the NC.\cite{Goupalov98,slms}
The short-range exchange contribution is almost independent 
of the NC shape and can result in an anisotropic splitting 
only in the case of 
low-symmetry phases.
This contribution is rather challenging to calculate.\cite{Dvorak13}

\begin{figure}[tbp]
  \centering{
   \includegraphics[width=0.65\linewidth]{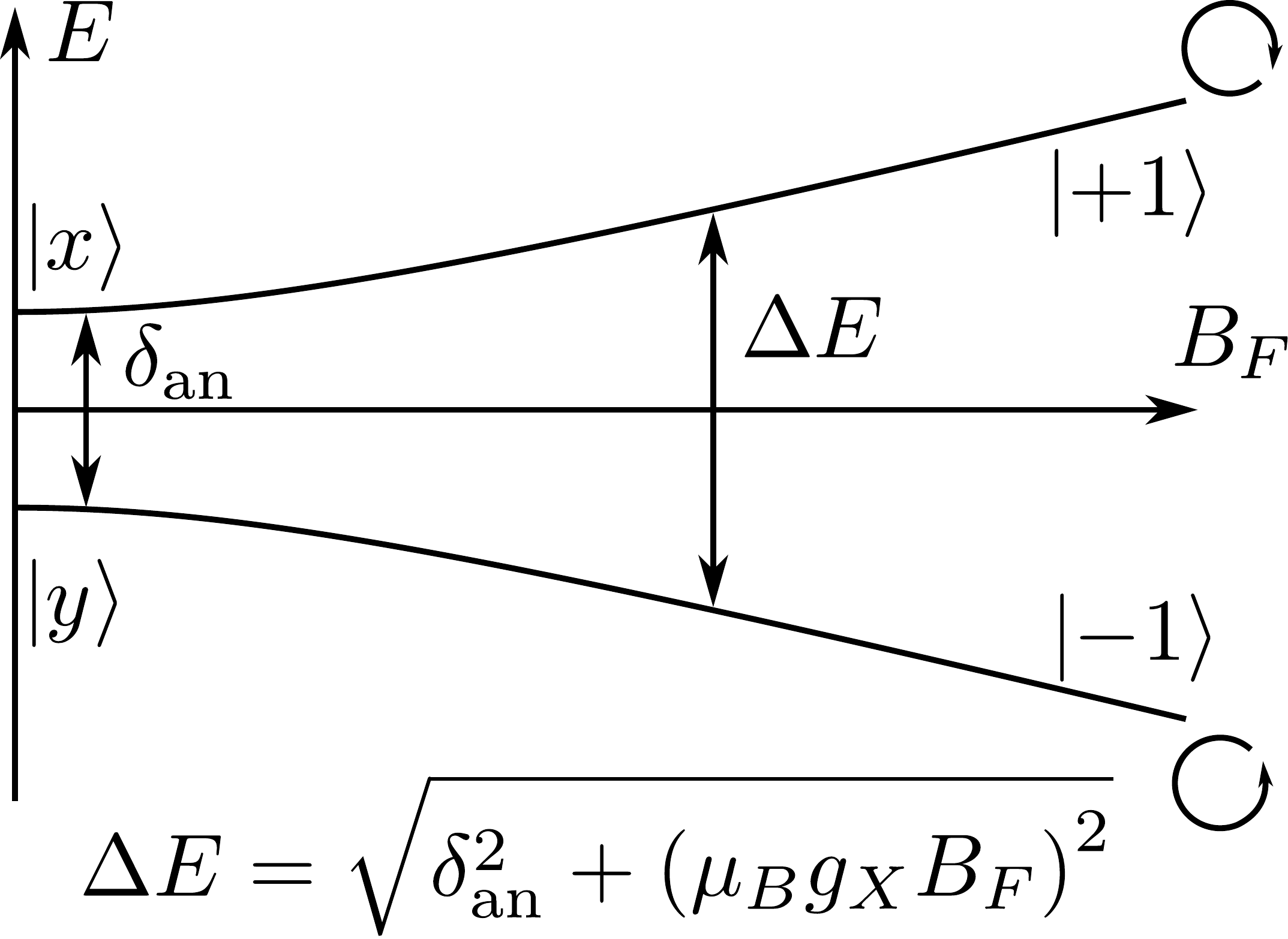}
  }
  \caption{Scheme of the $x$ and $y$ energy levels in magnetic field 
   $B_F\parallel z$. 
  } 
\label{fig:scheme_mf}
\end{figure}

In a magnetic field ($B_{F}\parallel z$), the structure of exciton levels changes.
If the field is oriented along one of eigenaxes of the 
structure, then the Zeeman splitting mixes the in-plane 
states, ``switching'' from $\ket{x}$, $\ket{y}$ states, enforced by the 
NC symmetry, to the circularly polarized $\ket{+1}$, $\ket{-1}$
states, see Fig.~\ref{fig:scheme_mf}.

\subsection{Short-range electron-hole exchange interaction in quantum dots: symmetry analysis}

The effective Hamiltonian describing short-range electron-hole exchange interaction for the exciton in a quantum dot having
$D_{2h}$ point symmetry takes the form
\begin{equation}
\Delta \hat{H}^{SR}_{exch}=I_x \, \hat{\sigma_x}^e \, \hat{\sigma_x}^h +
I_y \, \hat{\sigma_y}^e \, \hat{\sigma_y}^h + I_z \, \hat{\sigma_z}^e \, \hat{\sigma_z}^h \,,
\label{Hspin}
\end{equation}
where $\hat{\sigma_{\alpha}}^e$ and $\hat{\sigma_{\alpha}}^h$ are the Pauli 
matrices in the bases of the electron and hole spin states, respectively, 
and $I_x$, $I_y$, and $I_z$ are the exchange constants. 
For $D_{4h}$ point symmetry $I_x=I_y$. For $O_h$ point 
symmetry $I_x=I_y=I_z$. The eigenenergies of this effective Hamiltonian are
\begin{subequations}
\begin{align}
E_x& =-I_x+I_y+I_z \,, \label{Ex}
\\ 
E_y&=I_x-I_y+I_z \,, \label{Ey}
\\
E_z&=I_x+I_y-I_z \,,
\\
E_{dark}&=-I_x-I_y-I_z \,.
\end{align}
\end{subequations}

For the effective spin Hamiltonian~(\ref{Hspin}), the states polarized
in the plane perpendicular to the $z$ axis are decoupled from the dark 
state and the state polarized along $z$. Magnetic field $B$ applied 
along the $z$ axis mixes the states~(\ref{Ex}) and~(\ref{Ey}), and their
energies become
\begin{equation}
E_{\pm}={I}_z \pm \sqrt{\left(g_X \, \mu_B \, B \right)^2/4 + \left(I_x-I_y \right)^2} \,.
\end{equation}
The zero-field anisotropic splitting is $\delta_{\mathrm{an}}^{SR}=2\cdot|I_x-I_y|$.

When all the energy levels $E_x$, $E_y$, and $E_z$ are distinct, they can be characterized by the two non-negative values of splittings. In what follows it is convenient to denote the largest and the smallest of these values by $\delta_1$ and $\delta_2$, respectively. 
Then the remaining splitting is given by $\delta_1-\delta_2$ (see Fig.~\ref{fig:levels_scheme}).

Note that, if the splitting of exciton levels induced by the long-range 
electron-hole exchange interaction  is taken into account, then the exciton
fine structure can still be described by the effective spin 
Hamiltonian~(\ref{Hspin}) but with renormalized exchange constants 
$\tilde{I}_x$, $\tilde{I}_y$, and $\tilde{I}_z$. 
While $I_x=I_y$ for the short-range electron-hole exchange interaction 
in a quantum dot with $D_{4h}$ or $O_h$ crystal symmetry, the  
splittings, induced by the 
long-range electron-hole exchange interaction, result in the 
renormalized exchange constants $\tilde{I}_x \neq \tilde{I}_y$ if 
the sizes of the NC along $x$ and $y$ 
are different. 

\subsection{Long-range electron-hole exchange interaction in bulk perovskites 
and quantum dots: \texorpdfstring{\KP}{kp}\ analysis} \label{subsection:lr}

Within the effective mass approximation, the theory of the long-range electron-hole
exchange interaction in excitons in bulk semiconductors was put forward by
Pikus and Bir\cite{Pikus71,BirPikus} and by Denisov and
Makarov.\cite{Denisov73} For excitons confined in nanostructures the
long-range electron-hole exchange interaction was studied in
Refs.~\citenum{slms,Goupalov98,jcg,Goupalov00,Goupalov01,Goupalov03}.
Here we will start with the expression for the matrix element of the 
long-range electron-hole exchange interaction\cite{bookIvchenko}
\begin{multline}
\langle c, m', {\bf k}_c';v, n', {\bf k}_h'|V_{exch}^{\mathrm{LR}}|
c, m, {\bf k}_c;v, n, {\bf k}_h \rangle = \\
V^{-1} \, \frac{4 \pi e^2 \hbar^2}{\varepsilon_{\infty} m_0^2 E_g^2} \, \frac{({\bf Kp}_{m'\bar{n}'})({\bf Kp}_{m\bar{n}})^*}{K^2} \, \delta_{{\bf k}_c+{\bf k}_h,{\bf k}_c'+{\bf k}_h'} \,,
\end{multline}
where $V$ is the normalization volume, ${\bf K}={\bf k}_c+{\bf k}_h$ is the exciton wave vector, $\varepsilon_{\infty}$ is the dielectric permittivity on the frequency of the excitonic resonance, $e$ is the electron charge, $m_0$ is the free electron mass, $E_g$ is the band gap, ${\bf p}_{m\bar{n}}$ is the momentum interband matrix element between the electron states $m$ and $\bar{n}$, 
where $m$, $n$ enumerates bands, and the hole state $n$ and the electron 
state $\bar{n}$ are related by the time reversal operation.

In the basis of the exciton states polarized along
the axes $\alpha=x,y,z$ the matrix element of the 
long-range electron-hole exchange interaction 
in a crystal of the
$D_{4h}$ point symmetry takes the form
\begin{multline}
\langle \alpha, {\bf K} | V^{\mathrm{LR}}_{exch}| \beta, {\bf K}' \rangle= \\
V^{-1} \, \delta_{{\bf K},{\bf K}'} \, \frac{8 \, \pi \, e^2 \, \hbar^2}{\varepsilon_{\infty} \, m_0^2 \, E_g^2} \, \Xi_{\alpha \beta}({\bf K}) \,,
\label{LRmatrelem}
\end{multline}
where 
\begin{equation}\label{Xi1}
\begin{split}
& \hat{\Xi}({\bf K}) = \frac{\mathscr{K}^T\cdot\mathscr{K}}{K^2}
 \,,
\\
\mathscr{K} &= \left[ K_x p_{\perp} \frac{\cos{\xi}}{\sqrt2}, 
K_y p_{\perp}  \frac{\cos{\xi}}{\sqrt2}, K_z p_{\parallel} \sin{\xi} \right]
\,.
\end{split}
\end{equation}
Here $p_{\perp}=\langle S |\hat{p}_x|X \rangle=\langle S |\hat{p}_y|Y \rangle$ 
and $p_{\parallel}=\langle S |\hat{p}_z|Z \rangle$ are the Kane interband momentum 
matrix elements. 

In the cubic system (groups $T_d$ or $O_h$)
$\cos{\xi}=\frac{\sqrt{2}}{\sqrt{3}}$, $\sin{\xi}=\frac{1}{\sqrt{3}}$
(see Eqs.~(\ref{Bloch1},\ref{Bloch2})), $p_{\perp}=p_{||}=p_{cv}$,
and the matrix element~(\ref{LRmatrelem}) becomes
\begin{equation}\label{omegaLT}
\langle \alpha, {\bf K} | V^{\mathrm{LR}}_{exch}| \beta, {\bf K}' \rangle
=\delta_{{\bf K},{\bf K}'} \frac{\pi a_B^3}{V} \, \hbar \omega_{LT} \, \frac{K_{\alpha} \, K_{\beta}}{K^2}
\,,
\end{equation}
where $a_B$ is the bulk exciton Bohr radius,  
\begin{equation}
\hbar {\omega}_{LT}=\frac{8 \, e^2 \, \hbar^2 \, \, p_{cv}^2}{3 \, \varepsilon_{\infty} \, m_0^2 \, E_g^2 \, a_B^3} 
\end{equation}
is the longitudinal-transverse
splitting (cf. Ref.~\citenum{Goupalov00}).

For an exciton confined in a quantum dot with $O_h$ crystal structure 
which has the shape of a cuboid,
the resonant frequency renormalization of the 
$\alpha$-polarized confined exciton due to the 
long-range (non-analytic) electron-hole exchange interaction is given by\cite{slms}
\begin{equation}
\delta \omega^{(\alpha)}_0 =
\omega_{LT} \,\frac{\pi \, a^3_B}{V} \sum_{{\bf K}} \:
\frac{K_{\alpha}^2}{K^2} F^2({\bf K}) \,.
\label{resfreq2}
\end{equation}
where $F({\bf K})$ is the Fourier transform of the exciton envelope 
function with coinciding electron and hole coordinates $\Psi({\bf R}, {\bf R})$.
Here we choose the axes $\alpha=x,y,z$ along the principal axes 
of the cuboid.
Eq.~(\ref{resfreq2}) may be considered as a result of averaging of the 
expression~(\ref{omegaLT})
over the exciton wave vector ${\bf K}$.
Note that if the crystal structure has the lower symmetry, then in \eqref{resfreq2} 
${\omega}_{LT}$ should be renormalized to reflect the anisotropy 
of interband momentum matrix elements in the bulk. 

In what follows we will model the envelope wave function of a particle
(electron, hole or exciton) confined in the anisotropic quantum dot
by the Gaussian function
\begin{equation}\label{Gaussian}
\psi({\bf r})=\frac{\exp{\left( -\frac{x^2}{2 L_x^2}-\frac{y^2}{2 L_y^2}-\frac{z^2}{2 L^2}
\right)}
}{\pi^{3/4} \, \sqrt{L_x L_y L}} \,,
\end{equation}
where $2 L_x$, $2 L_y$, and $2 L$ are, respectively, the quantum dot sizes along the $x$, $y$, 
and $z$ directions. 

One can distinguish two distinct regimes of exciton confinement in a quantum dot. 
In the strong confinement regime, when the size of the NC is small
compared with the exciton Bohr radius, 
\begin{equation}\label{eq:LR_weak}
\Psi({\bf R},{\bf R})=\psi_e({\bf R}) \psi_h({\bf R})\;.
\end{equation}
In the weak confinement regime, when the NC size is large compared 
with the exciton Bohr radius, the exciton 
is localized within the NC as a whole and 
\begin{equation}\label{eq:LR_strong}
\Psi({\bf R},{\bf R})=\frac{\psi_{exc}({\bf R})}{\sqrt{\pi a_B^3}}\;.
\end{equation}

\begin{table}[bp]
\begin{tabular}{|c|ccc|}
\hline
Material                         &  CsPbCl\textsubscript{3} & CsPbBr\textsubscript{3} & CsPbI\textsubscript{3} \\ \hline
$E_g$, eV                        &  2.82  & 2.00 & 1.44 \\
$\varepsilon_{\infty}$           &  4.07  & 4.96 & 6.32 \\
$a_{B}$, \AA                       &  25.0  & 35.0 & 60.0 \\
$\frac{\hbar}{m_0} \, p_{cv}$, eV $\cdot$ \AA &  11.9  & 11.9 & 11.5 \\
$\hbar\omega_{LT}$, meV          & 10.7   & 6.33 & 1.79 \\ \hline
\end{tabular}
\caption{\KP\ parameters used in calculations.}\label{tbl:params}
\end{table}

Substituting these functions into Eq.~(\ref{resfreq2}) 
and assuming weak shape anisotropy 
($|L_x-L| \ll L$, $|L_y-L| \ll L$) one 
obtains for the fine anisotropic splitting 
$\delta_{\mathrm{an}}^{LR} = \hbar \left( \delta \omega^{(x)}_0-\delta \omega^{(y)}_0 \right)$
\begin{equation}
\delta_{\mathrm{an}}^{LR}= \frac{\hbar {\omega}_{LT}}{5 \, \sqrt{2 \pi}} \,
\left( \frac{a_B}{L} \right)^3 \, \frac{L_y-L_x}{L}
\label{strongresult}
\end{equation}
in the strong confinement regime and
\begin{equation}
\delta_{\mathrm{an}}^{LR} = \frac{2 \, \hbar {\omega}_{LT}}{5} \,
\frac{L_y-L_x}{L}
\label{weakresult}
\end{equation}
in the weak confinement regime.
Note that, in the orthorhombic phase, the resonance frequencies
are renormalized due to anisotropy of the interband momentum matrix element
and the splitting is non-zero 
even if the envelope function is isotropic ($L_x=L_y=L$).

\begin{figure}[tbp]
  \centering{
   \includegraphics[width=0.9\linewidth]{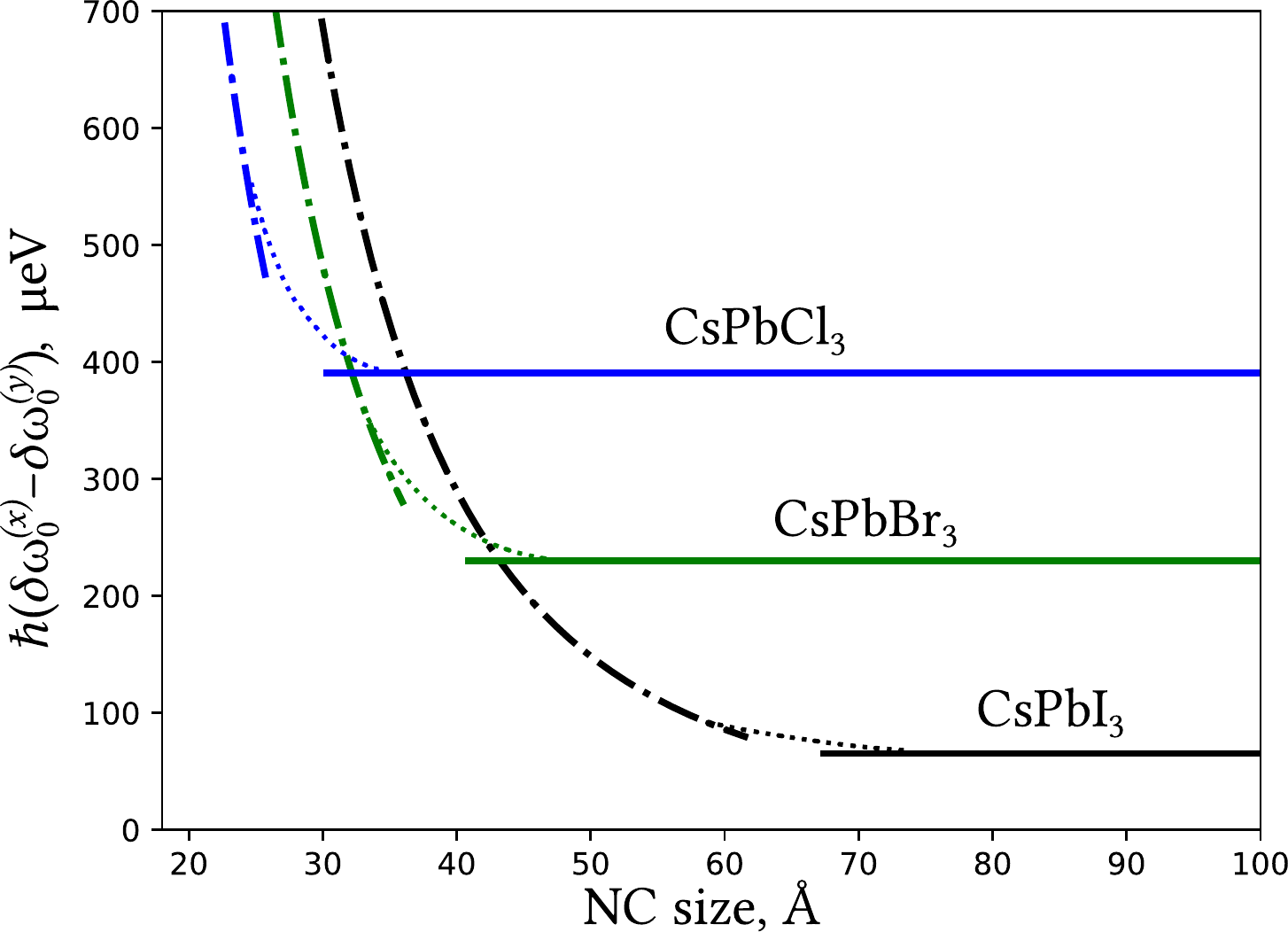}
  }
 \caption{Bright exciton splitting due to NC anisotropy 
  calculated in \KP\ approximation for three 
  perovskite materials as a function of NC size 
  assuming the 10\% shape anisotropy of the NC.
  The \KP\ parameters are given in Table~\ref{tbl:params}.
  Solid lines show the weak confinement regime \eqref{weakresult},
  the dash-dotted lines the strong confinement 
  regime \eqref{strongresult} and the dashed lines are 
  the guides to the eye.
  } 
\label{fig:split_all}
\end{figure}
\begin{figure*}[btp]
  \centering{
   \includegraphics[width=0.35\linewidth]{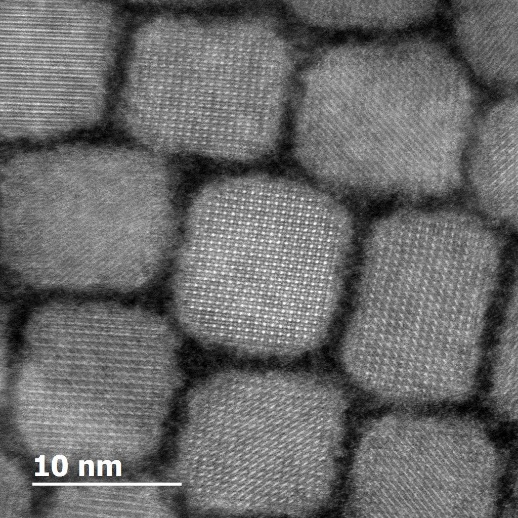}\hfil%
   \includegraphics[width=0.35\linewidth]{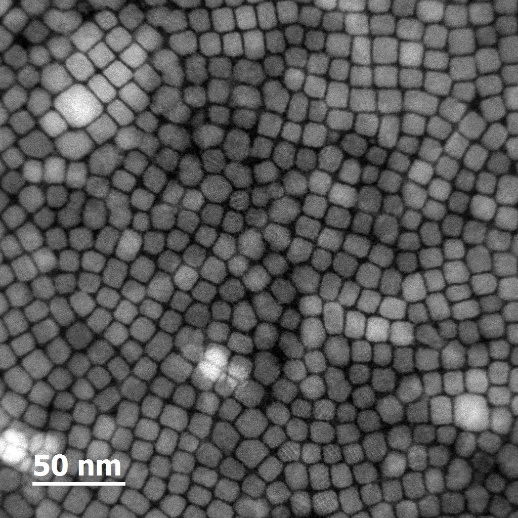}
  }
  \caption{ Atomic resolution ADF-STEM image of 
    CsPbI\textsubscript{3} NCs studied here.
    The size of the NCs $\simeq$10~nm. 
  } 
\label{fig:TEM}
\end{figure*}
To estimate the value of $\delta_{\mathrm{an}}^{LR}$ we
use the parameters of bulk perovskite 
CsPbI\textsubscript{3}, CsPbBr\textsubscript{3}, 
and CsPbCl\textsubscript{3}. These parameters, taken from Ref.~\citenum{Protesescu15}, are summarized in 
Table~\ref{tbl:params}. The value of the
interband momentum matrix element is estimated from the 
electron and hole masses assuming negligible contribution 
from the remote bands.
From these parameters, using Eqs.~(\ref{strongresult},\ref{weakresult}), we calculated the value of 
the anisotropic splitting $\delta_{\mathrm{an}}^{LR}$
shown in Fig.~\ref{fig:split_all} as a function of the NC size, 
for a fixed NC in-plane shape anisotropy of 10\%.
For CsPbI\textsubscript{3} NCs with the size of $\sim10$~nm the anisotropic 
splitting due to the long-range electron-hole exchange interaction is 
approximately 65~$\mu$eV.

\section{Synthesis of C\lowercase{s}P\lowercase{b}I\textsubscript{3} nanocrystals}\label{sec:synthesis}

Cesium lead iodide perovskite NCs (CsPbI\textsubscript{3} NCs) 
were synthesized following the method of Protesescu et al.\cite{Protesescu15}
First, Cs-oleate was prepared by mixing 0.814~g of Cs\textsubscript{2}CO\textsubscript{3} with 40~mL of octadecene (ODE) and 2.5~mL of oleic acid (OA), with all reactants dried at 120$^\circ$C for 1~h. The mixture was stirred at 150$^\circ$C in an inert atmosphere until completion of the reaction. For the NCs formation, 5~mL of ODE and 0.188~mmol of lead (II) iodide (PbI\textsubscript{2}) were dried for 1~h at 120$^\circ$C in N\textsubscript{2} atmosphere. 
After the removal of moisture, 0.5~mL of dried OA  and 0.5~mL of dried oleylamine were added to the reaction flask, and the temperature
was raised up to 180$^\circ$C.
After complete solvation of the lead salt, 0.4~mL of the previously warmed up Cs-oleate solution  was injected. A few seconds later, the NCs solution was
quickly cooled down with an ice bath. The product was purified by
centrifugation and subsequently redispersed in hexane. The synthesized colloidal solution was drop-cast deposited on a quartz glass substrate and dried until a continuous thin film was formed.

The sample was investigated by a JEOL 2100F scanning transmission electron 
microscope (STEM) equipped with a delta corrector, which compensates for the aberration
up to the fifth order. Figure~\ref{fig:TEM} shows the annular dark-field (ADF) 
STEM image of the drop-cast deposited CsPbI\textsubscript{3} NC ensemble 
with atomic resolution, revealing its closely packed morphology.
From Fig.~\ref{fig:TEM} we conclude that the NCs have the shape of the 
cuboid with edge size close to 10~nm and the average aspect ratio about 
10\%. One of the main axis of all NCs is normal to the layer plane and the 
orientation of the main axes of different NCs in the layer plane
is random, so the layer as a whole is isotropic in the lateral
direction. This fact is reflected in the spectroscopic measurements, see below.

\section{PL measurements}\label{sec:PL}

\begin{figure}[tbp]
  \centering{
   \includegraphics[width=0.8\linewidth,trim={0 0.9cm 0 0}]{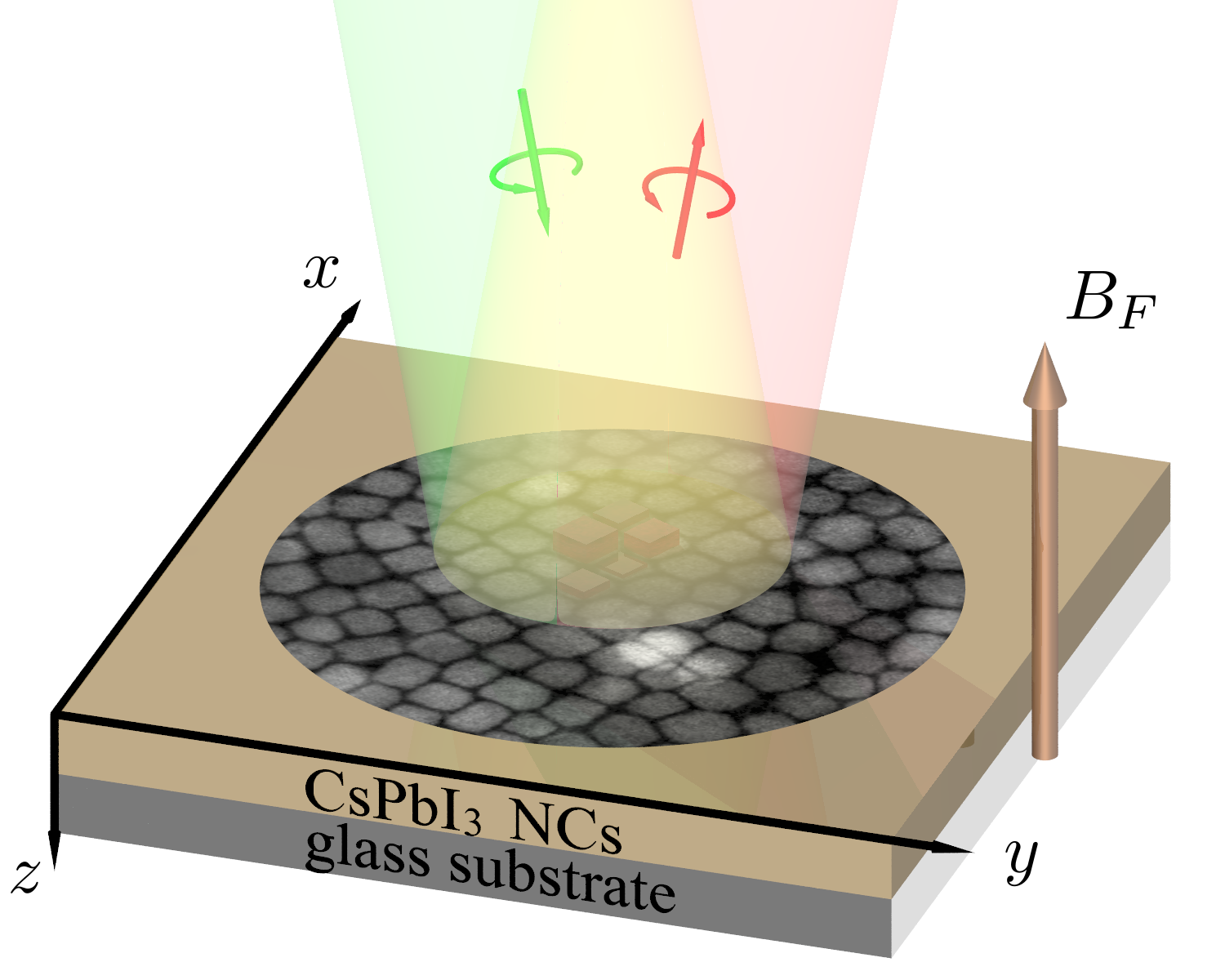}
   }
  \caption{ The geometry of the experiment.
  } 
\label{fig:expt_geom}
\end{figure}

The geometry of the experiment for studying polarized PL 
at cryogenic temperatures is shown in 
Fig.~\ref{fig:expt_geom}. 
PL was excited by a titanium-sapphire laser,
tunable in the range of 700--820 nm in continuos wave regime.
The beam of the excitation 
light was directed along the normal to the crystal surface ($z$-axis). 
The light spot diameter was $\simeq$150~$\mu$m, which is at least 
four orders of magnitude larger than typical NC size and guarantees 
averaging of the observed PL over the large number of 
NCs.
The PL was detected in the backscattering geometry at a small angle to the $z$-axis, and recorded by the Horiba iHR-550 spectrometer and an avalanche photodiode (APD). 
The APD was connected to a scheme of the two-channel photon counting detector from which
data was transferred to a computer. The sample was subjected to a longitudinal magnetic field $\textbf{B}_F \parallel z$ (Faraday geometry) produced by a resistive magnet.
We note that the PL intensity
considerably rises with the decrease of temperature
from room temperature to the helium temperatures, 
contrary to e.g. Si NCs.\cite{Heitmann04}

The experimental technique follows the work by Dzhioev et al.~\cite{Dzhioev97}. 
Polarized luminescence is determined completely by specifying four
components of the Stokes parameters \cite{bookBlum81}, which give the following 
information: 
(a) full light intensity $I$; 
(b) the degree of circular polarization;
(c) the degree of linear polarization with respect to the pair of orthogonal axes ($x$, $y$);
(d) the degree of linear polarization 
with respect to the axes ($x'$, $y'$) rotated 
by an angle of $45^\circ$ 
relative to the ($x$, $y$) axes
around the $z$-axis.

We are interested only in the Stokes parameters related to the 
polarization of light. It should be borne in mind that in addition
to the three Stokes parameters characterizing the polarization of 
the secondary radiation, it is also necessary to have full information 
on the polarization of the excitation light, which, in turn, is also 
characterized by three Stokes components. Thus, in the general case, 
we have a set of $3\times 3=9$ different measurements, because for 
each of the three polarizations of the excitation light it is 
possible to measure three Stokes parameters of the secondary 
radiation:
\begin{equation}\label{eq:stokes}
\rho_c^{\alpha}       = \frac{I_{\sigma^+}^{\alpha} - I_{\sigma^-}^{\alpha}}{I_{\sigma^+}^{\alpha} + I_{\sigma^-}^{\alpha}},\;
\rho_{\ell}^{\alpha}  = \frac{I_{x}^{\alpha} - I_{y}^{\alpha}}{I_{x}^{\alpha} + I_{y}^{\alpha}},\;
\rho_{\ell'}^{\alpha} = \frac{I_{x'}^{\alpha} - I_{y'}^{\alpha}}{I_{x'}^{\alpha} + I_{y'}^{\alpha}},\;
\end{equation}
where the upper index $\alpha$ indicates the polarization of the excitation light, 
which is set by a fixed polarizer in the excitation channel for 
right-handed circular 
polarization ($\alpha=\sigma^+$), linear polarization along $x$ ($\alpha=L$) 
and $x'$ ($\alpha=L'$) axes. Lower index refers to the registration channel,
in which the analyzer modulates the circular polarization 
from $\sigma^+$ to $\sigma^-$ ($\rho_c^{\alpha}$),
the linear polarization from the $x$- to the $y$-axis ($\rho_{\ell}^{\alpha}$),
or the linear polarization from the $x'$- to the $y'$-axis ($\rho_{\ell'}^{\alpha}$).
The total PL intensity does not depend on the choice of the 
orthogonal components: 
$I=I_{\sigma^+}^{\alpha}+I_{\sigma^-}^{\alpha} = I_{x}^{\alpha}+I_{y}^{\alpha} = I_{x'}^{\alpha}+I_{y'}^{\alpha}$.

However, instead of measuring the polarization degrees $\rho_c^{\alpha}$, $\rho_{\ell}^{\alpha}$ and $\rho_{\ell'}^{\alpha}$ defined in accordance with  \eqref{eq:stokes}, we applied a modulation technique where the analyzer is in a fixed position and the sample is pumped by the incident light changing its polarization from a circular or linear to the orthogonal at a frequency of 42 kHz. In this case, the setup measures the \textit{effective} Stokes parameters

\begin{equation}\label{eq:stokes2}
  \rho^c_{\alpha}       = \frac{I^{\sigma^+}_{\alpha} - I^{\sigma^-}_{\alpha}}{I^{\sigma^+}_{\alpha} + I^{\sigma^-}_{\alpha}},\;
  \rho^{\ell}_{\alpha}  = \frac{I^{x}_{\alpha} - I^{y}_{\alpha}}{I^{x}_{\alpha} + I^{y}_{\alpha}},\;
  \rho^{\ell'}_{\alpha} = \frac{I^{x'}_{\alpha} - I^{y'}_{\alpha}}{I^{x'}_{\alpha} + I^{y'}_{\alpha}},\;
\end{equation}
where the lower index fixes the position of the analyzer and the top index refers to the
excitation channel, whose polarization is modulated.
This technique allows one to avoid the effects of 
dynamic nuclear polarization. Strictly speaking, the polarization 
parameters, defined by the formulas \eqref{eq:stokes} and 
\eqref{eq:stokes2}, are different. 
However, it is possible to show \cite{Dzhioev97_PRB} that if the effects of 
dichroism (circular or linear) are negligible, then equations 
\eqref{eq:stokes2} can be considered as the usual Stokes parameters,
which characterize circularly (or linearly) polarized luminescence 
in the case of polarized excitation.

An appearance of the circularly polarized excitonic
PL under the circularly polarized excitation ($\rho^c_{\sigma^+}\not= 0$) 
is known as the optical orientation of excitons,\cite{Bir72,bookPikus}
while the linear polarization of the excitonic PL under the 
linearly polarized excitation ($\rho_L^{\ell},\rho_{L'}^{\ell'}\not= 0$) 
is known as the optical alignment of excitons. 
An appearance of the circular (linear) polarization of the PL under 
the linear (circular) excitation is referred to as conversion of 
the orientation to alignment (alignment to orientation).

\section{Experimental results}\label{sec:results}

\begin{figure*}[tbp]
  \centering{
   \includegraphics[width=\linewidth]{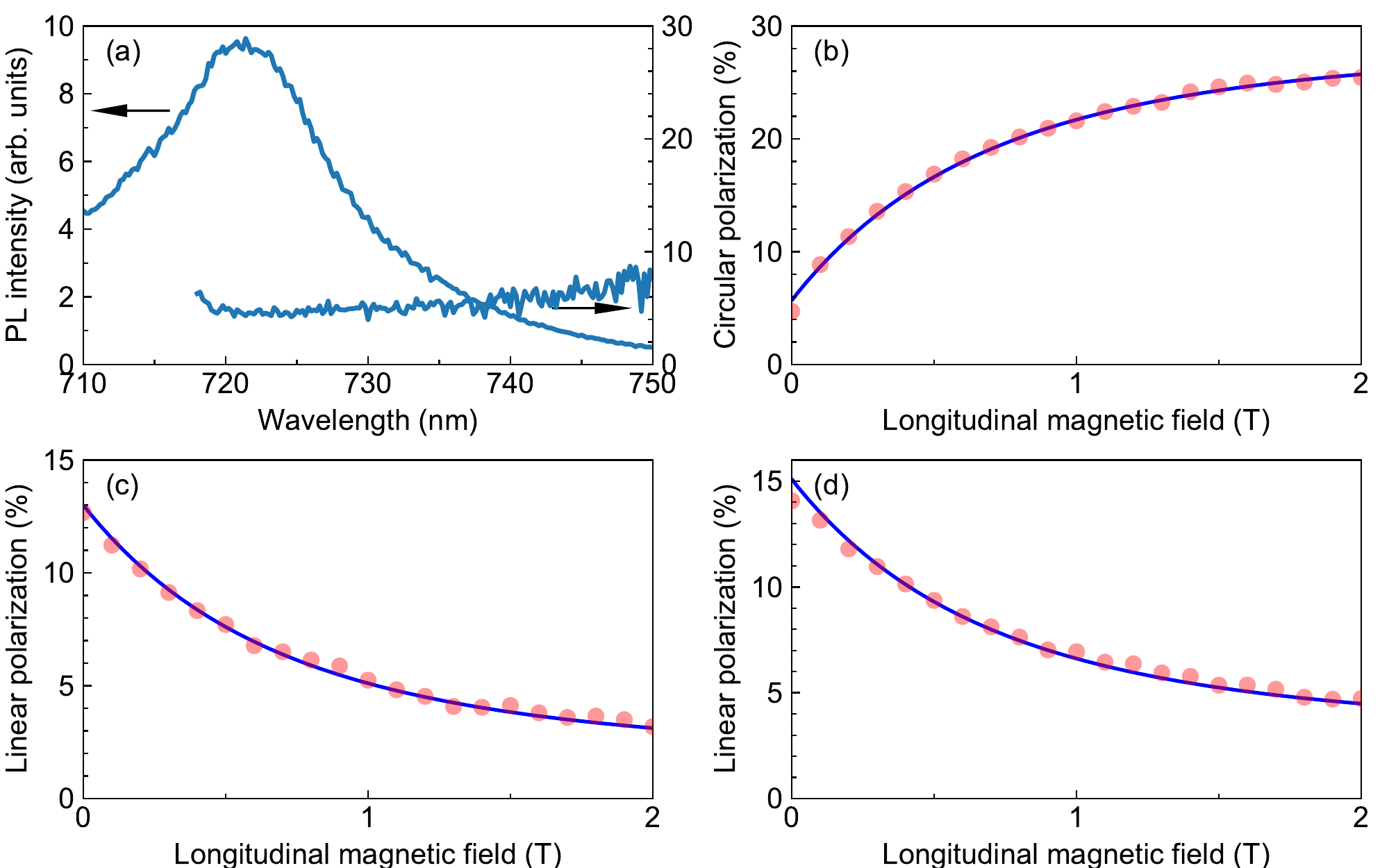}
  }
  \caption{ Polarization spectroscopy of the PL of 
  the sample at $\lambda_{\mathrm{ex}}=705$~nm,
  $W=0.6$~W/cm$^2$, $T = 2$~K. 
  (a) PL spectra and optical orientation at zero magnetic field.
  Symbols show the polarization measurements of the PL in the longitudinal field 
  detected at $\lambda_{\mathrm{det}}=721$~nm:
  (b) optical orientation $\rho^c_{\sigma^+}(B_F)$; 
  optical alignment $\rho_L^{\ell} (B_F)$ (c)  and (d) $\rho_{L'}^{\ell'} (B_F)$.
  In panels b,c,d the fit of the polarization after Eqs.~\eqref{eq:Pc}, \eqref{eq:Pl} 
  averaged over shape anisotropy distribution \eqref{eq:distr_LR} is 
  shown in solid lines, see text.
  } 
\label{fig:res}
\end{figure*}

Figure~\ref{fig:res}a shows the spectral dependencies of PL intensity, $I(\lambda)$, and PL optical orientation, $\rho^c_{\sigma^+}(\lambda)$, in zero magnetic field 
at the excitation wavelength $\lambda_{\mathrm{ex}}=705$~nm, power density
$W=0.6$~W/cm$^2$, and temperature $T=2$~K. Polarization measurements in a 
longitudinal magnetic field were carried out at 
$\lambda_{\mathrm{det}}=721$~nm, which corresponds to the maximum of the PL
intensity. An external magnetic field, $B_F$, restores the optical 
orientation (increases the degree of circular polarization), see 
Fig.~\ref{fig:res}b. 
An optical alignment of $\rho_L^{\ell} =13$\% is observed in PL when exciting with light linearly polarized along the $x$-axis in 
zero field,
see Fig.~\ref{fig:res}c. 
The observed optical alignment is virtually isotropic: when illuminated by light 
polarized along the $x'$, the polarization degree is $\rho_{L'}^{\ell'} =14$\%,
see Fig.~\ref{fig:res}d. This indicates a random distribution of dipoles over the ensemble of 
NCs. The degree of linear polarization decreases 
with increasing magnetic field. 
This indicates that the phenomena of optical orientation and alignment in this system have the same physical nature. 

The suppression of the optical orientation at zero magnetic field, together with the strong optical alignment of excitons indicates 
that the bright exciton state is split at zero field into linearly polarized components.
It is worth to note that there is a non-zero
polarization $\rho^c_{\sigma^+}=4$\%
at zero magnetic field. 
In a transverse magnetic field polarization decreases in a characteristic field 
of 7~mT. 
This contribution may originate from the excitons in 
cubic phase NCs, single electrons, or trions.
This will be considered in more details in another work, 
but the small residual circular polarization and 
strong optical alignment indicate that, in our experiments, 
the trions which are known to dominate the optical properties 
of similar systems\cite{Canneson17} 
do not affect our results significantly.

Note that the 
transformation (conversion) of one polarization of excitons into another 
(for example, the appearance of linear polarization under circularly polarized pumping) 
is absent. Alignment of excitons is the hallmark of the anisotropy of the system, 
either microscopical or mesoscopical.
The absence of polarization conversion shows that there are almost no
NCs with degenerate bright exciton states.

The optical orientation experiments reveal the 
structure of exciton levels and 
allow one to measure the magnitude of the splitting of the exciton bright 
state, $\delta_{\mathrm{an}}$, averaged over an ensemble of NCs, 
even in the absence of a sufficient spectral resolution. In a weak magnetic field  ($B_{F}\ll\delta_{\mathrm{an}}/\mu_B g_{X}\equiv b$, where $g_{X}$ is the exciton $g$-factor, $\mu_B$ is the Bohr magneton) the circular polarization of PL is absent because the $\sigma^{\pm}$-light excites a coherent superposition of the $x$- and $y$-states split by $\delta_{\mathrm{an}}\simeq 120$~$\mu$eV.
This splitting causes fast beats between the two eigenstates which average circular polarization out during the exciton lifetime $\tau\sim0.2$ ns ($\tau\gg \hbar / \delta_{\mathrm{an}} \sim 10$ ps).\cite{Fu17} When the magnetic field increases ($B_{F} \geq b$),  the degree of the PL circular polarization also increases, thus restoring optical orientation of the excitons. The optical orientation is completely restored when $B_{F} \gg b$. In turn, the optical alignment of the excitons already exists in zero magnetic field when excitation light is polarized along the $x$ or $y$ axis. Magnetic field $B_{F} > b$ converts $\ket{x/y}$ states into $\ket{\pm 1}$ states decreasing the linear polarization of the PL,
as illustrated in Fig.~\ref{fig:scheme_mf}.

Here we do not explicitly consider the role of the dark exciton 
and the $z$-polarized states of the bright excitons. 
When the dark exciton states have energy close to the bright ones, 
one may observe the resonance in the PL polarization intensity 
due to the anticrossong between dark and bright excitons.\cite{IvchenkoKaminskii95, Gourdon98}
In our experiments, this anticrossing is not present and we conclude that
the possible presence of the dark excitons does not affect the 
polarized PL signal.
The presence of $z$-polarized bright excitons may not affect the 
circular polarization profile, though it may in principle affect 
its amplitude. The $z$-polarized bright exciton state may contribute 
to the linear polarization 
$\rho^{\ell}_{L}(B_F)$, $\rho^{\ell'}_{L'}(B_F)$.
However, as long as all three polarizations in Fig.~~\ref{fig:res}~b,c,d 
are perfectly fitted using the same set of parameters, we conclude that 
in our experiments $z$-polarized bright excitons do not contribute to 
the linear polarization as well.

In order to describe the PL polarization measurements, 
one can introduce the exchange-related frequencies 
$\Omega_x$ and $\Omega_y$ 
as the coefficients in the effective 
Hamiltonian for the radiative doublet $\ket{\pm1}$\cite{Dzhioev97}
\begin{equation}
  \mathcal{H}_{\mathrm{eff}}
  = \frac{1}2 \begin{pmatrix}
    \mu_B g_X B_F & \hbar\Omega_x - i\hbar\Omega_y \\
    \hbar\Omega_x + i\hbar\Omega_y & -\mu_B g_X B_F
  \end{pmatrix}\,.
\end{equation}
In the following we neglect the exciton spin relaxation.

To average over the ensemble, one has to assume the distribution function 
of the exchange splitting $\delta_{\mathrm{an}}$. 
When the short-range exchange contribution to the splitting is small compared with the 
long-range exchange one, the splitting is proportional to the shape anisotropy,
see Eq.~(\ref{eq:LR_weak}, \ref{eq:LR_strong}).
The distribution of the shape anisotropy is typically Gaussian:\cite{Elazzouzi08}
\begin{equation}\label{eq:distr_LR}
\mathcal{P}(\Omega) = 
\frac1{\sigma_{\delta}\sqrt{2\pi}} \exp\left[ -\frac{\Omega^2}{2\sigma_{\delta}^2} \right]\;,
\end{equation}
where $\sigma_{\delta}$ is the dispersion of the long range
exchange splitting and the amplitude of the splitting for 
given NC is $\Omega$.
Note that the orientation of the NC principal axes in the layer is 
random, which results in i.e., $\langle \Omega_x \Omega_y \rangle =0$

The averaging over the lateral distribution of the 
orientation of NCs 
is easier to consider separately: assuming the 
NCs with splitting of linearly polarized states $\Omega$ with the 
random distribution of principal
axes in the substrate plane
and assuming the exciton lifetime in the radiative states $\ket{ \pm 1}$,
$\tau\gg1/\Omega$, one 
can obtain for the PL circular polarization under resonant 
circularly-polarized excitation in the longitudinal 
magnetic field\cite{Dzhioev97,bookIvchenko}
\begin{equation}\label{eq:Pc}
  P_c(B_z;\Omega) = P_c (0) \frac{ \left(\mu_B g_X B_F\right)^2}{ \left( \mu_B g_X B_F\right)^2 + \hbar^2\Omega^2}\:.
\end{equation}
Similar consideration gives\footnote{We 
give only one linear polarizaion because in the isotropic system
$P_{\ell}(B_z;\Omega)=P_{\ell'}(B_z;\Omega)$.}
for the linear polarization\cite{Dzhioev97,bookIvchenko}
\begin{equation}\label{eq:Pl}
  P_{\ell}(B_z;\Omega) = P_{\ell} (0) \frac{  \hbar^2\Omega^2 }{ \left( \mu_B g_X B_F\right)^2 + \hbar^2\Omega^2}\:.
\end{equation}

The polarization for the ensemble of different NCs is then obtained 
by averaging of Eqs.~(\ref{eq:Pc}, \ref{eq:Pl}) over the distribution of 
the exciton splitting \eqref{eq:distr_LR}:
\begin{equation}\label{eq:Paver}
  P_{\xi}(B_z) = \int \mathcal{P}(\Omega) P_{\xi}(B_z;\Omega) d\Omega
\end{equation}

In Fig.~\ref{fig:res}~b,c,d we show the results of the 
fit of the optical orientation and alignment experimental measurements
using Eqs.~(\ref{eq:distr_LR}-\ref{eq:Paver})
assuming the long-range exchange 
origin of the splitting
caused by the NC shape 
\eqref{eq:distr_LR} (solid line).
In fitting procedure we assume that there is additional  
polarization caused by other mechanisms: presence of trions for 
circular polarization and dichroism for the linear polarization. 
Both the amplitude of the polarization $P_{\xi}(0)$ and
the shifts $\Delta P_{\xi}$ are considered as 
free parameters.\footnote{For completeness, the fit in 
Fig.~\ref{fig:res}b,c,d is given for the following values:
$\Delta P_{c}=5.7$, $\Delta P_{\ell}=1.7$, $\Delta P_{\ell'}=2.9$, 
$P_{c}(0)=23.6$, $P_{\ell}(0)=11.6$, $P_{\ell'}(0)=12.5$.}
From the best fit we extract 
the dispersion of the long-range exchange splitting $\sigma_{\delta}$.

The best fit assuming Gaussian distribution of the bright exciton splittings \eqref{eq:distr_LR}
gives an excellent agreement with experimental data for 
$\hbar\sigma_{\delta}/\mu_B g_X = 0.917$~T.
Using the exciton $g$-factor extracted from Raman data\footnote{V.F. Sapega, unpublished}
 $g_X\simeq 2.3$ (which is close to typical value in similar 
systems, see Ref.~\onlinecite{Fu17})
we may extract the value of 
the dispersion of the long-range exchange splitting
$\hbar\sigma_{\delta} = 122$~$\mu$eV. 
In experiment, we have the Gaussian distribution of NC shape anisotropy with the 
standard deviation 10\%, see Fig.~\ref{fig:TEM}. 
This means that the optical orientation may be explained if we 
assume that for 10\% shape anisotropy we have the bright exciton
splitting 122~$\mu$eV and this splitting is proportional to the 
shape anisotropy.

There is also a possibility to explain the 
optical orientation and alignment assuming that the splitting of 
the bright excitons originates from the 
short-range exchange splitting 
$\delta_{\mathrm{an}}^{SR}$ and the long-range exchange 
is negligible. 
The NCs are located on the substrate randomly, with one of the 
principal axes 
of each NC normal to the substrate plane, which means that for each NC 
the splitting is equally distributed between 
three possibilities: $\delta_1^{SR}$, $\delta_2^{SR}$, or 
$\delta_1^{SR}-\delta_2^{SR}$.
The results of polarization spectroscopy in Fig.~\ref{fig:res}
may also be quite well fitted under this assumption 
with the crystal-field splittings of bright excitons 
$\delta_1^{SR}=123$~$\mu$eV and $\delta_2^{SR}=14$~$\mu$eV.

\section{Discussion}\label{sec:discussion}

The value of the bright exciton fine structure splitting, averaged over an ensemble
of CsPbI\textsubscript{3} NCs, measured in our experiment is close to $120$~$\mu$eV.
This value exceeds our theoretical estimate of $65$~$\mu$eV for the bright exciton splitting resulting from the NC shape anisotropy of 10~\% and caused by the long-range electron-hole exchange interaction. However, it is
$1.5$ to $5$ times less than the values measured on individual NCs in 
Ref.~\citenum{Yin17}, although the mean size of the NCs was about the same.
Moreover, Yin {\it et al.} emphasized that they also investigated some samples, where the fine structure splittings for all NCs, on which they performed single-dot measurements, were below their spectral resolution of $200$~$\mu$eV, although the same synthesis procedure was nominally adopted for all the samples.\cite{Yin17} Provided that the cesium lead halide perovskites are known to feature different crystal phases and that co-existence of these phases in nanostructures has been reported,\cite{Alivisatos16} one can speculate that, under some conditions, which are difficult to control, the NCs can have different phases or be inhomogeneous, which would affect their exciton fine-structure splittings.

Indeed, in addition to the long-range electron-hole exchange 
interaction, which is sensitive to the NC shape anisotropy, there is its short-range counterpart
sensitive to the underlying crystal phase.
In the cubic phase, the anisotropic part of the short-range (analytic)
electron-hole exchange interaction is zero. In the tetragonal phase, the
additional anisotropy of the short-range (analytic) electron-hole exchange
interaction does not necessarily affect the splitting between the $x$- and $y$-polarized
exciton states ($\delta_2^{SR}=0$), while our estimates show that the long-range (non-analytic)
exchange interaction is comparable with the observed anisotropic splitting.
In the orthorhombic phase, 
however, the long-range (non-analytic) electron-hole exchange interaction
has an additional in-plane anisotropy, and there may be a significant 
contribution from the short-range electron-hole exchange interaction 
to the anisotropic splitting. 

Note that, if we assume tetragonal phase of the NCs, then 
taking into account only the short-range part of the exchange 
interaction cannot explain our results.
The polarization dependencies of Fig.~\ref{fig:res} cannot be fitted 
if one takes $\delta_2^{SR}=0$.

\section{Conclusion}\label{sec:conclusions}
In conclusion, we have presented measurements of optical orientation 
and alignment of excitons
in ensembles of CsPbI\textsubscript{3} NCs at 
cryogenic temperatures. 
From our experiment we 
conclude that there is an anisotropic splitting of 
bright exciton levels in NCs.
The experimental data may be fitted if one assumes that the 
splitting is 
related to the NC shape anisotropy and amounts 122~$\mu$eV. Our theoretical estimate, 
basing on the not-well-known value of the interband momentum matrix element for CsPbI$_3$ and 
assuming 10\% NC shape anisotropy, extracted from the TEM measurements, gives about half of this value.
We note, however, that both the anisotropic shape of NCs and the 
possible low-symmetry phase of the underlying crystal structure, 
as well as a combination thereof,  may cause the anisotropic 
fine-structure splitting; optical spectroscopy alone cannot be 
used to rule out either of these possibilities.
Further investigations are called for to unambiguously determine 
the crystal phase of cesium lead halide NCs at cryogenic temperatures 
and to determine the value of the interband momentum matrix element for
CsPbX$_3$ (X=I, Br, Cl).

\section*{Acknowledgments}
The authors acknowledge fruitful discussions with E.L.~Ivchenko and M.M.~Glazov.
The work of MON was  
supported by the Government of the Russian
Federation (contract \#14.W03.31.0011 at the Ioffe Institute).
The work of SVG was supported by
the National Science Foundation (NSF-CREST Grant HRD-1547754).
Ch.dW, L.G. and T.G. acknowledge financial support by NWO 
(Nederlandse organisatie voor Wetenschappelijk Onderzoek) 
and Y.F. and T.G. thank Osaka University for International Joint Research Promotion Program.
J.L. and K.S. acknowledge JST-ACCEL and JSPS KAKENHI (JP16H06333 and P16823)
The work of LBM was supported by LETI Personal Grant for Scientific Projects of Young Researchers
The work of INY was partially supported by Program of 
Russian Academy of Sciences.

\bibliography{oo_in_perovskites}

\end{document}